\documentclass{article}
\usepackage[utf8]{inputenc}
\usepackage{authblk}
\usepackage{algorithmic}
\usepackage[lined,boxed,commentsnumbered, ruled,vlined,linesnumbered, french]{algorithm2e}              
\usepackage{booktabs}
\usepackage{setspace}
\usepackage[margin=1.25in]{geometry}
\usepackage{graphicx}
\graphicspath{ {./figures/} }
\usepackage{subcaption}
\usepackage{caption}
\usepackage{float} 

\usepackage{subcaption}
\usepackage{amsmath}
\usepackage{hyperref}
\usepackage{listings}

\title{Approximating Euler’s Totient Function Using Linear Regression on RSA Moduli}

\author[1*$\dag$]{Gilda Rech BANSIMBA}
\author[2$\dag$]{Regis F. BABINDAMANA}
\author[3$\dag$]{Beni Blaug N. IBARA}

\affil[1,2,3]{Department of Mathematics and Computer Science, Université Marien Ngouabi, Faculté des Sciences et Techniques, BP: 69, Brazzaville, Congo}

\date{}

\begin{document}

\maketitle

\begin{abstract}
The security of the RSA cryptosystem is based on the intractability of computing Euler's totient function $\phi(n)$ for large integers~$n$. Although deriving $\phi(n)$ deterministically remains computationally infeasible for cryptographically relevant bit lengths, and machine learning presents a promising alternative for constructing efficient approximations.
In this work, we explore a machine learning approach to approximate Euler’s totient function $\phi$ using linear regression models. We consider a dataset of RSA moduli of $64$, $128$, $256$, $512$ and $1024$ bits along with their corresponding totient values. The regression model is trained to capture the relationship between the modulus and its totient, and tested on unseen samples to evaluate its prediction accuracy. Preliminary results suggest that $\phi$ can be approximated within a small relative error margin, which may be sufficient to aid in certain classes of RSA attacks. This research opens a direction for integrating statistical learning techniques into cryptanalysis, providing insights into the feasibility of attacking cryptosystems using approximation-based strategies.
\end{abstract}

\section{INTRODUCTION}
The integration of machine learning (ML) and cryptography has become a key research focus, enabling novel advancements in both securing and compromising cryptographic systems. Unlike classical cryptography, which depends on mathematical hardness assumptions, ML especially deep and reinforcement learning—provides data-driven methods for automated cryptanalysis, cipher optimization, and vulnerability detection \cite{alabaichi2023,pandey2024}. 
It is the case in cryptanalysis with Neural distinguishers where Deep learning models have broken reduced-round variants of lightweight ciphers (e.g., SPECK, SIMON) by identifying non-random statistical patterns \cite{gohr2019}; with Generative adversarial attacks where GANs have been used to reconstruct secret keys from partial side-channel data \cite{timon2019} or with Profiled side-channel attacks where Supervised learning (e.g., Random Forests, CNNs) outperforms traditional methods like template attacks in key extraction \cite{benadjila2020}. Herein,
The Euler’s totient function, denoted by $\phi$, is a function that counts numbers less than $n$ relatively prime to $n$. This function plays a key role in number theory and modern cryptographic systems, especially in RSA encryption \cite{rsa}. In the RSA scheme, the security relies on the computational hardness of factoring large semiprime numbers and on the inaccessibility of $\phi (n)$ for a given public modulus $n=p\times q$, where $p$ and $q$ are large prime numbers. However, gaining even partial information about $\phi$ can potentially compromise RSA’s security. As related work, in \cite{sierpinski}, Sierpinski proved that $\phi(n)\leq n-\sqrt{n}$, while in \cite{Kendall}, Kendall and al. proved that for $n>30$, $\phi(n)>n^{2/3}$ and in \cite{Hatalova} Hatalova and al. proved for $n\geq 3$, that $\phi(n)>\frac{\log{}2}{2}\frac{n}{\log{}n}$. Recently in \cite{Fang} Fang and al. showed a lower bound for $n\geq 3$, $\phi(n)\geq \frac{n}{\exp^{\gamma}loglogn}+O(\frac{n}{(\log{}\log{}n)^{2}})$, with $\gamma$ the Euler-Mascherone constant. 
In this paper, we propose a linear regression-based approach to approximate $\phi$ for RSA moduli of various sizes, including 64, 128, 256, 512, and 1024 bits. The main objective of this study is to investigate whether it is possible to predict $\phi$ with sufficient accuracy to initiate attacks on RSA, particularly those relying on approximate values of the totient function.

The motivation behind this research lies in the intersection between machine learning and cryptanalysis. By training a regression model on known pairs of semiprimes and their totients, we aim to provide an empirical tool that could support cryptanalytic techniques in cases where direct factorization is infeasible. As results of this study, it outcomes that for a given RSA modulus $n$ of size $s$-bits a good lower bound approximation is 
\begin{eqnarray*}
n_{s}-2\alpha_{s}+2<\phi(n_{s}) \ \ \text{with} \ \
\alpha_{s}=
\left\lbrace
\begin{array}{ll}
1637177340 \ \ \ \text{if} \ \ s= 64 \ bits\\ \ \\
75557863700000000704512 \ \ \ \text{if} \ \ s= 128 \ bits\\ \ \\
1928325650000000078912455744507\\3532909932502488404174072446976 \ \ \ \text{if} \ \ s= 256 \ bits\\ \ \\
1339709139999999990480967267719955950833\\
1598925069799668366025204959482723407083 \ \ \ \text{if} \ \ s= 512 \ bits\\
4864298075302130999709681365436895620871\\
68500607188390117376 
\end{array} \right.
\end{eqnarray*}
The intercept of our linear regression function is dependent on the size of the target modulus input whereas the slope remains invariant and the model built using a paremeter related to $\phi(n)$ introduced by Bansimba and al. in \cite{GR} from results on hyperbolas \cite{Gilda, Rech}.

\section{DATA}
The security of RSA cryptosystems fundamentally depends on the computational hardness of factoring large integers and computing Euler's totient function $\phi(n)$. To investigate our potential approximation methods, we generate a comprehensive dataset of RSA moduli across cryptographic bit lengths. Using the Rabin-Miller primality test, we construct $1$ million prime pairs for each of five bit sizes ($32$ to $512$ bits), yielding moduli $n = pq$ ranging from $64$ to $1024$ bits—a spectrum covering both legacy and modern security parameters.

In the RSA setting, the public parameters are the modulus $n$ and the public encryption exponent $e$. We build our predictors for $\phi$ focusing on known features, mainly the public exponent $n$.\/\  
Additionally, from \cite{GR}, for a given RSA modulus, the Hyper-X $X=2(n-\varepsilon)(n+1)-(n-1)^2$ and Hyper-Y $Y=4n(n-\varepsilon)^2$ and the Euler's totient function is expressed in terms of $\varepsilon$ as $\phi_{\varepsilon}(n)=2(\varepsilon+1)$ with $\varepsilon$ an odd integer. We therefore consider $\varepsilon$ as our target from which we derive an approximation of $\phi$.\\
Each of our 5 datasets has a shape of \emph{(1000000, \ 4)} with the following features \emph{prime p}, \emph{prime q}, \emph{product n} and \emph{$\varepsilon$}.\\
The total size of the generated data on disk is $1362.3 $ Mb $\ \approx \ 1.362$ GB.\\
\begin{figure}[H]
  \centering
  
  \begin{subfigure}[b]{0.45\textwidth}
    \centering
    \includegraphics[width=\textwidth]{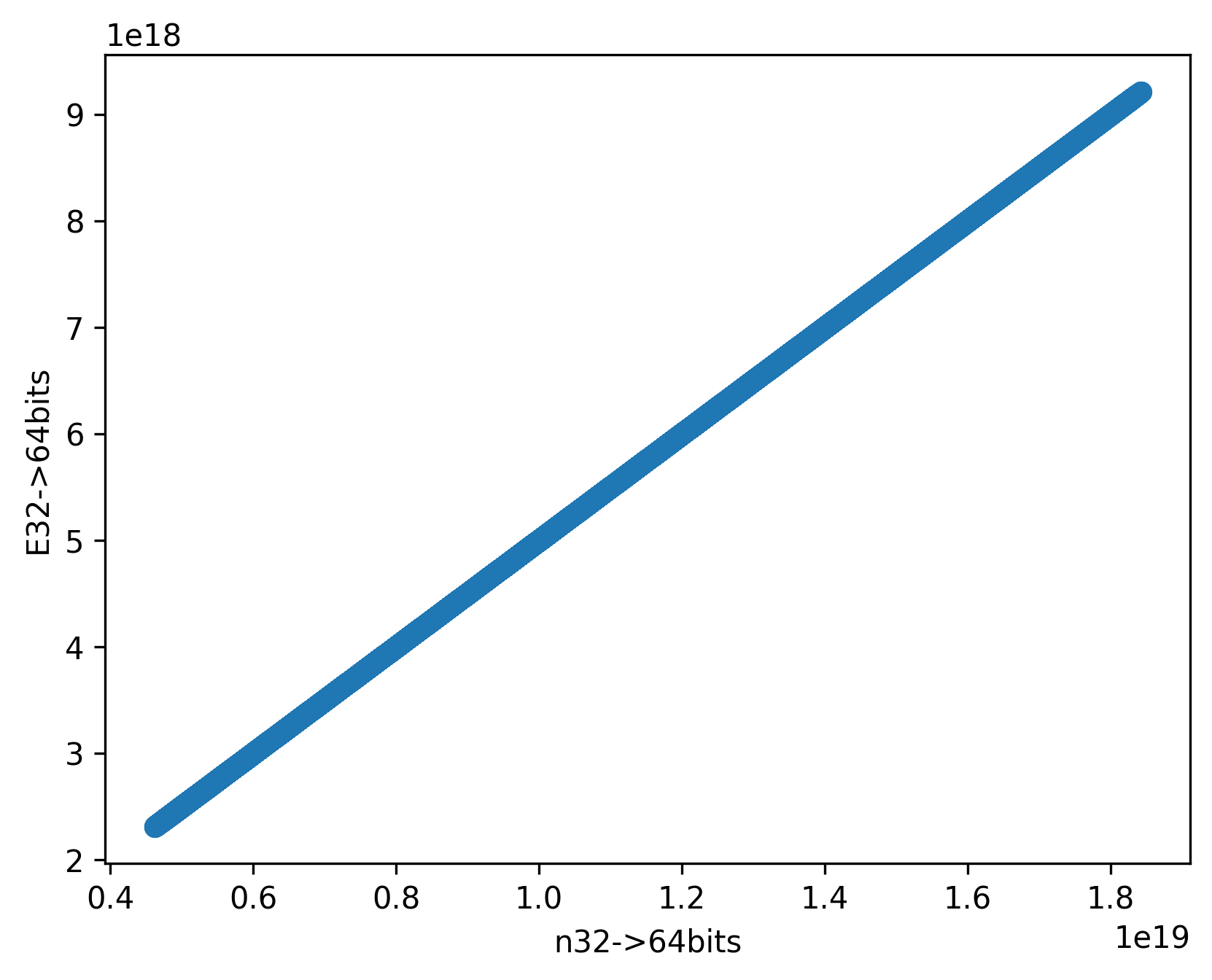}
    \caption{64-bit modulus}
  \end{subfigure}
  \hfill
  \begin{subfigure}[b]{0.450\textwidth}
    \centering
    \includegraphics[width=\textwidth]{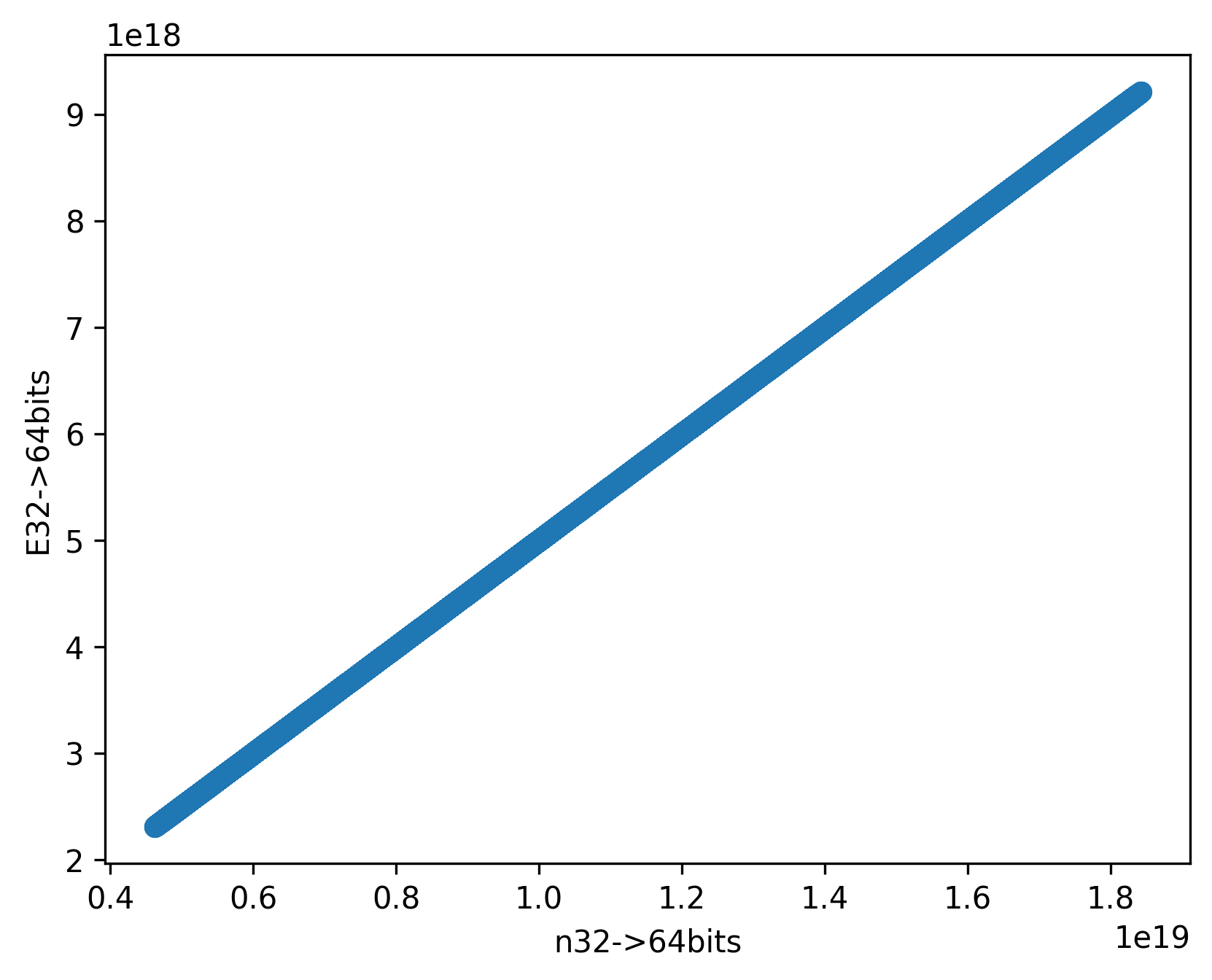}
    \caption{128-bit modulus}
  \end{subfigure}
  
  \vspace{0.5cm}
  
  \begin{subfigure}[b]{0.450\textwidth}
    \centering
    \includegraphics[width=\textwidth]{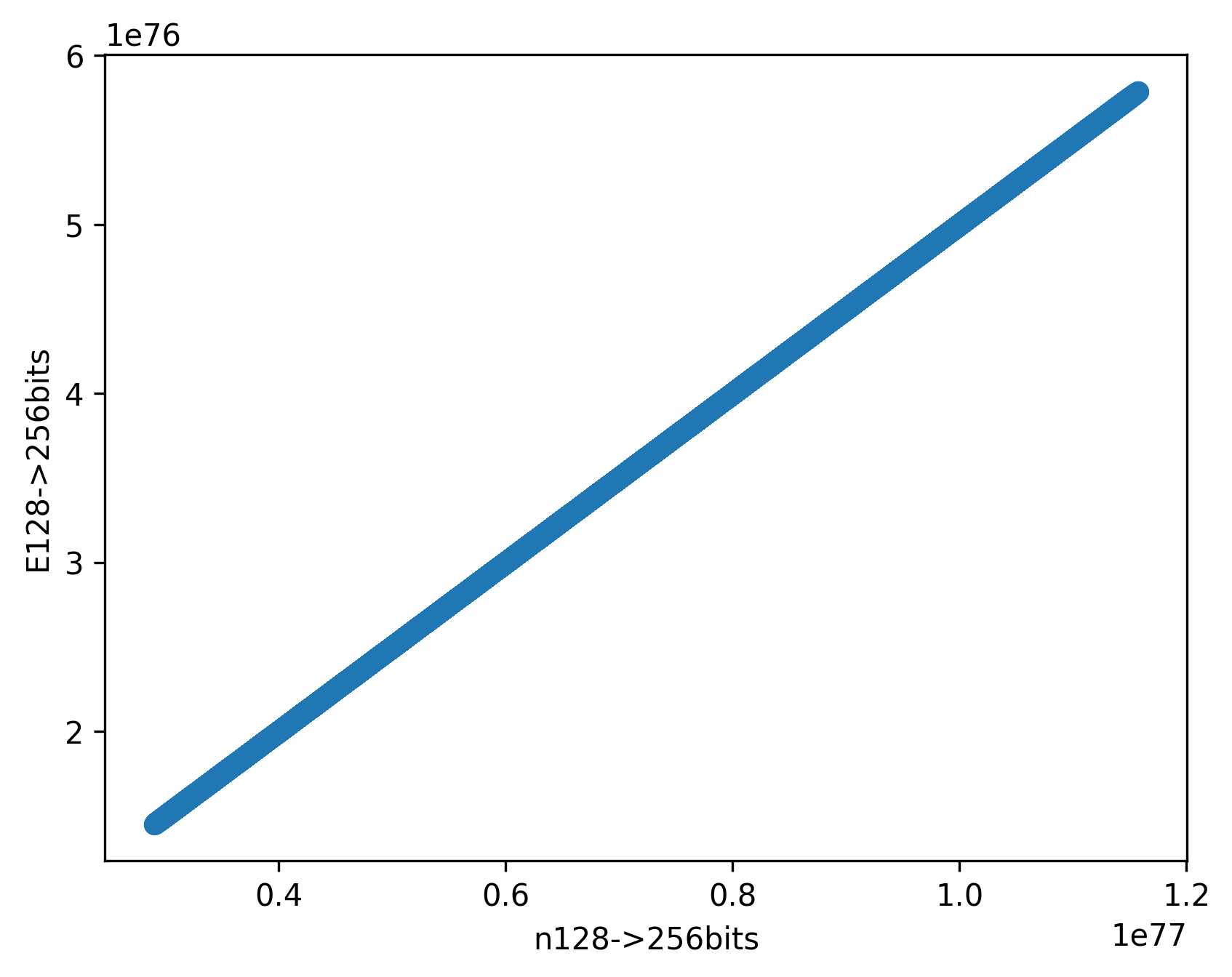}
    \caption{256-bits modulus}
  \end{subfigure}
  \hfill
  \begin{subfigure}[b]{0.450\textwidth}
    \centering
    \includegraphics[width=\textwidth]{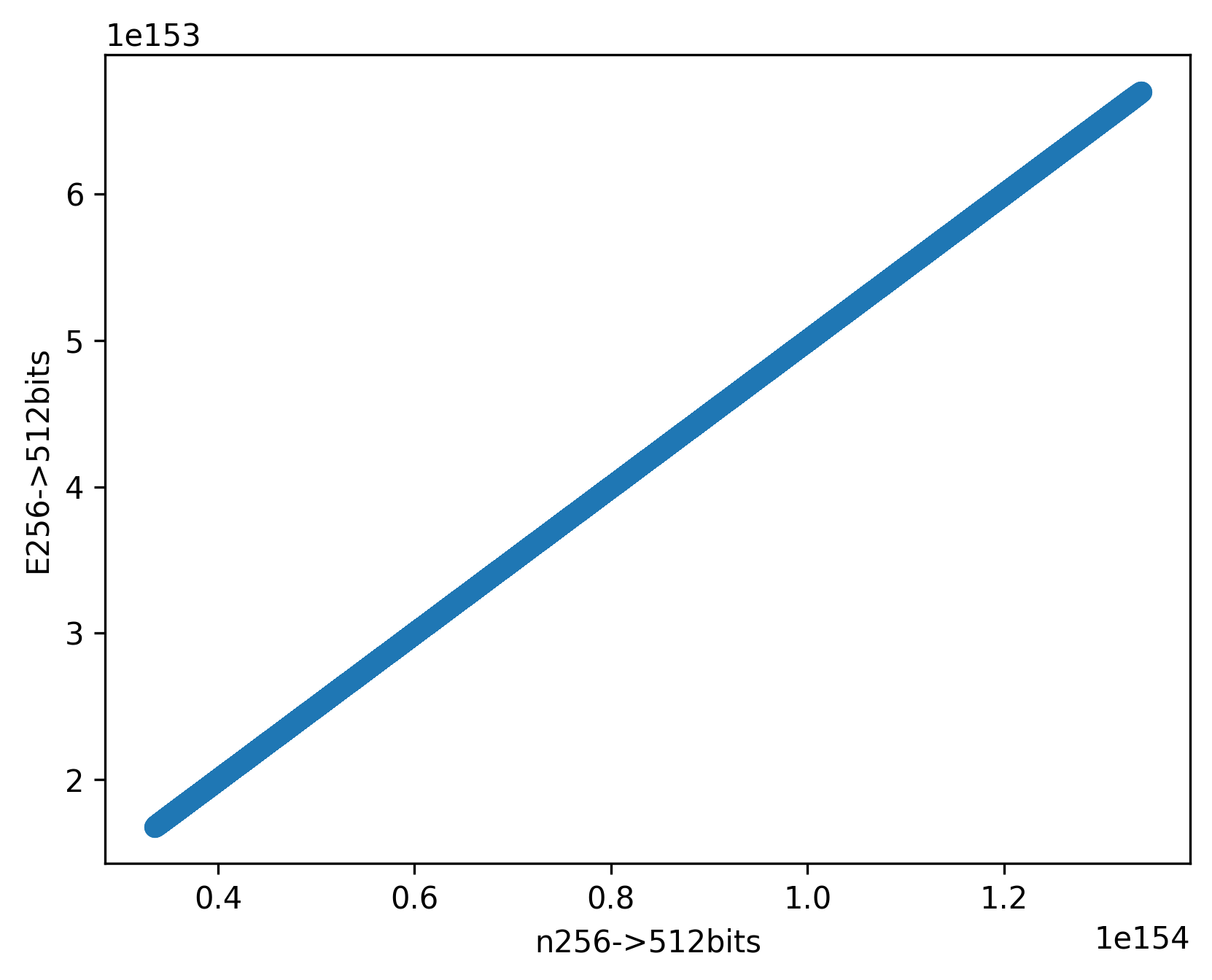}
    \caption{512-buts modulus}
  \end{subfigure}
  
  \caption{linear relation}
\end{figure}

\section{Linear regression}
The choice of this predictor is motivated by the linear relationship between the modulus $n$ and $\varepsilon$ as illustrated in the previous figures from our observations. We target $\varepsilon$ considering the public parameter $n$.
Training: We consider $80\%$ of data for training and $20\%$ for test.
We get the following predictors\\ 
$\displaystyle\varepsilon_{pred}=\frac{1}{2}n_{64bits}-1637177340 \ \ \ \ < \ \ \ \varepsilon_{true}$\\
$\displaystyle\varepsilon_{pred}=\frac{1}{2}n_{128bits}-75557863700000000704512 \ \ \ \ < \ \ \varepsilon_{true}$\\
$\displaystyle\varepsilon_{pred}=\frac{1}{2}n_{256bits}-19283256500000000789124557445073532909932502488404174072446976 \ < \ \varepsilon_{true}$\\
$\displaystyle\varepsilon_{pred}=\frac{1}{2}n_{512bits}-13397091399999999904809672677199559508331598925069799668366025\\
204959482723407083486429807530213099970968136543689562087168500607188390117376 \ \ \ \ \ \ \ < \ \ \ \varepsilon_{true}$\\
These predictors predict correctly the first digits of $\varepsilon$ with an accuracy of\\ $0.9999999999999999$. \\
From \cite{GR}, since $\phi_{\varepsilon}(n)=2(\varepsilon+1)$, then $\varepsilon=\frac{1}{2}\phi_{\varepsilon}-1$, and finally for a modulus $n$ of size $s$-bits, we have the following good lower bound $n_{s}-2\alpha_{s}+2<\phi(n_{s})$, where: 

\begin{eqnarray*}
\alpha_{s}=
\left\lbrace
\begin{array}{ll}
1637177340 \ \ \ \text{if} \ \ s= 64 \ bits\\ \ \\
755578637000\\
00000704512 \ \ \ \text{if} \ \ s= 128 \ bits\\ 
\end{array} \right. ;
\left\lbrace
\begin{array}{ll}
1928325650000000078912455744507\\3532909932502488404174072446976 \ \ \ \text{if} \ \ s= 256 \ bits\\ \ \\
13397091399999999904809672677199559\\
50833159892506979966836602520495948 \ \ \ \text{if} \ \ s= 512 \ bits\\
27234070834864298075302130999709681\\
36543689562087168500607188390117376 
\end{array} \right.
\end{eqnarray*}

\section{Performance Evaluation}
To evaluate the performance of the linear regression model, we calculated the following metrics:
\begin{itemize}
    \item[$\bullet$] \textbf{Mean Absolute Error (MAE)}: Measures the average absolute difference between the predicted and true values.
    \item[$\bullet$] \textbf{Mean Squared Error (MSE)}: Measures the average of the squared differences between predicted and actual values.
    \item[$\bullet$] \textbf{$R^2$ Score}: Indicates the proportion of the variance in the target variable that is predictable from the features.
\end{itemize}
The model achieved excellent performance on the test data, with the following results:
\subsection*{For $64$-bit RSA moduli}
\begin{itemize}
    \item[$\bullet$] \textbf{Mean Absolute Error}  \begin{lstlisting}[basicstyle=\ttfamily]
MAE = 44363744.
\end{lstlisting}

    \item[$\bullet$] \textbf{Mean Squared Error} \begin{lstlisting}[basicstyle=\ttfamily]
MSE = 3515104257251425.
\end{lstlisting}

    \item[$\bullet$] \textbf{$R^2$ Score}: $= 1$.
\end{itemize}

\subsection*{For $128$-bit RSA moduli}
\begin{itemize}
    \item[$\bullet$] \textbf{Mean Absolute Error} \begin{lstlisting}[basicstyle=\ttfamily]
MAE = 134592592141251222962176.
\end{lstlisting}
    
    \item[$\bullet$] \textbf{Mean Squared Error}: \begin{lstlisting}[basicstyle=\ttfamily]
MSE = 28827933808137106633233156038605117907112296448.
\end{lstlisting}
    
    \item[$\bullet$] \textbf{$R^2$ Score}: $= 1$.
\end{itemize}

\subsection*{For $256$-bit RSA moduli}
\begin{itemize}
    \item[$\bullet$] \textbf{Mean Absolute Error} \begin{lstlisting}[basicstyle=\ttfamily]
MAE = 86127861827171225812729173253369519324865652273211915280318464.
\end{lstlisting}
    \item[$\bullet$] \textbf{Mean Squared Error} \begin{lstlisting}[basicstyle=\ttfamily]
MSE = 109787140330933359805816487015567105532303473585912821903236
82791047413294171013888526774770254096252272992149477479818461184
\end{lstlisting}
    \item[$\bullet$] \textbf{$R^2$ Score}: $= 1$.
\end{itemize}

\subsection*{For $512$-bit RSA moduli}
\begin{itemize}
    \item[$\bullet$] \textbf{Mean Absolute Error} \begin{lstlisting}[basicstyle=\ttfamily]
MAE = 2158221180004875038593846411154319102619139020124252672438441
3881266429533645130831960003058584293891198831761932460309490975212
81629159424.
\end{lstlisting}

    \item[$\bullet$] \textbf{Mean Squared Error (MSE)}: \begin{lstlisting}[basicstyle=\ttfamily]
MSE = 7173312704232366570330070766040898910486467300816465983756463
3128010073756853677720215581184618517647688218224346291845386992436
4634287155208021539822465755901263677612897062675737431733258551871
4487345593317325586130783769940455055663930234100886462655357402764
640829797564416
\end{lstlisting}

    \item[$\bullet$] \textbf{$R^2$ Score}: $= 1$.
\end{itemize}

As shown, the model predicts $\varepsilon$ with high accuracy across all modulus sizes. The $R^2$ scores are consistently close to 1. indicating that the linear regression model explains most of the variance in $\varepsilon$ for each modulus size.

\section{ Graphical Analysis}
To visually assess the performance of the model, we plot the predicted values against the actual values for each modulus size. Figure 1 shows the scatter plots for the ($64$, $128$, $256$, $512$)-bit RSA modulus, where the blue points represent the true values of $E$ and the red line shows the predicted values from the linear regression model.

\begin{figure}[H]
  \centering
  
  \begin{subfigure}[b]{0.45\textwidth}
    \centering
    \includegraphics[width=\textwidth]{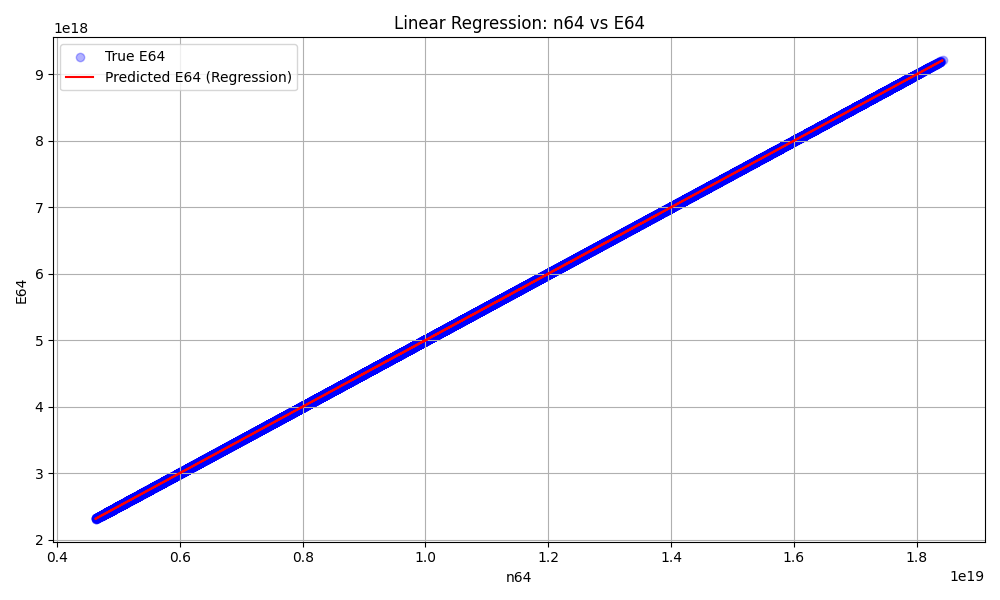}
    \caption{64-bit modulus}
  \end{subfigure}
  \hfill
  \begin{subfigure}[b]{0.450\textwidth}
    \centering
    \includegraphics[width=\textwidth]{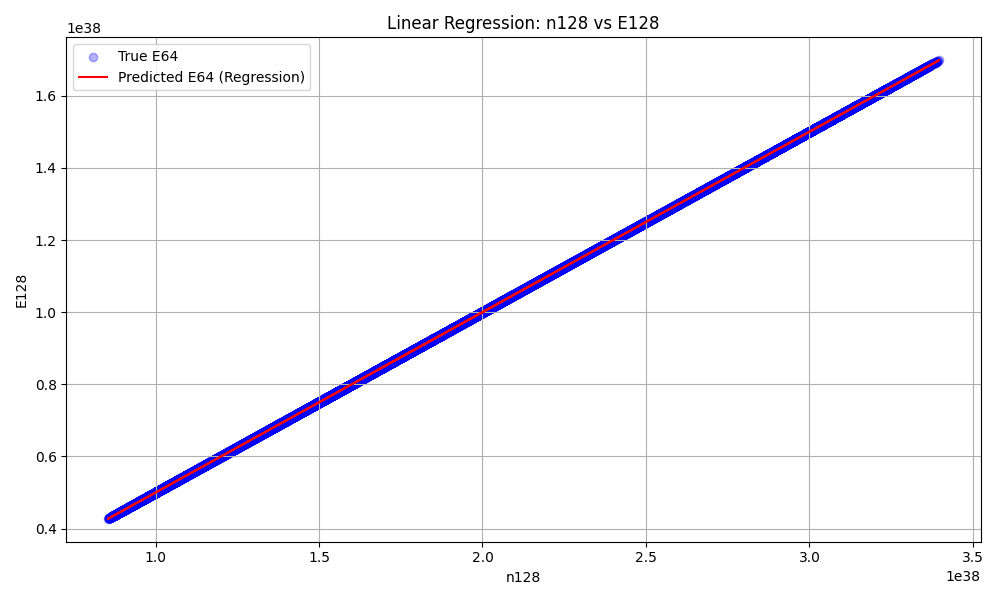}
    \caption{128-bit modulus}
  \end{subfigure}
  
  \vspace{0.5cm}
  
  \begin{subfigure}[b]{0.450\textwidth}
    \centering
    \includegraphics[width=\textwidth]{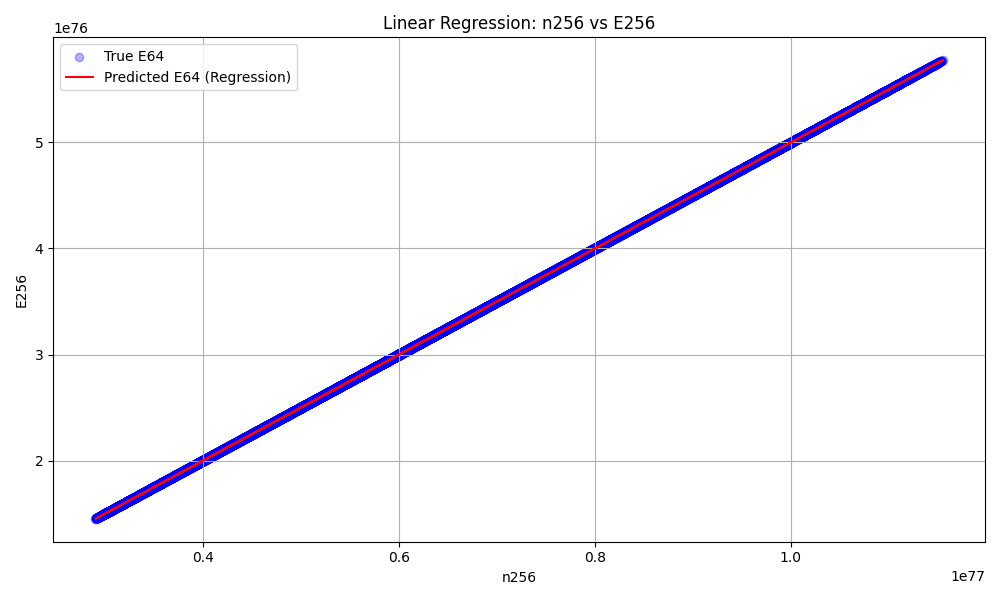}
    \caption{256-bits modulus}
  \end{subfigure}
  \hfill
  \begin{subfigure}[b]{0.450\textwidth}
    \centering
    \includegraphics[width=\textwidth]{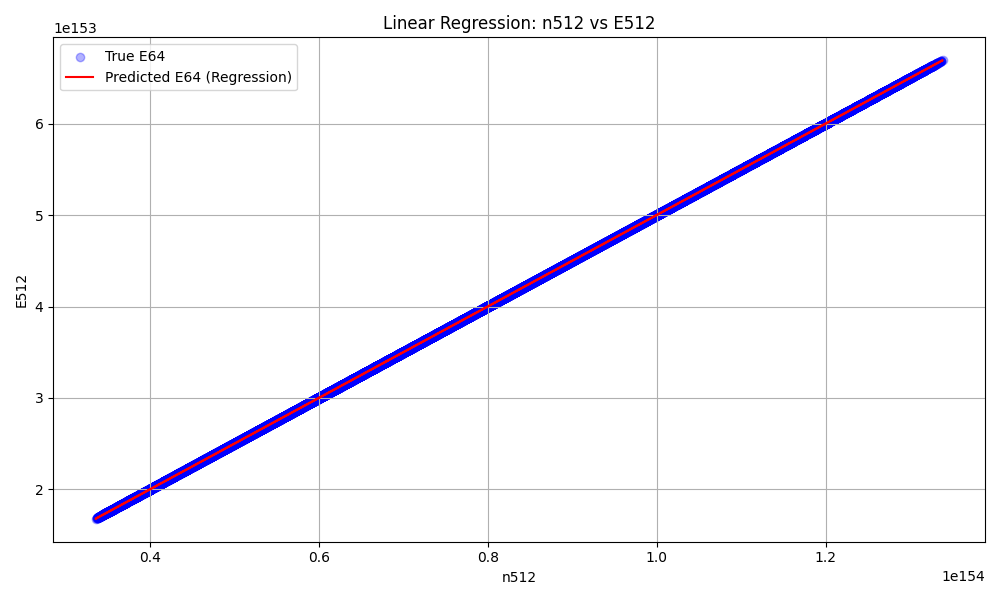}
    \caption{512-buts modulus}
  \end{subfigure}
  
  \caption{Performance of the model}
\end{figure}

\section{Error Distribution}
Figure 6 shows the distribution of errors (i.e., the difference between predicted and true values) for the $64$-bit RSA modulus. The errors are symmetrically distributed around zero, indicating that the model’s predictions are unbiased. This behavior is consistent across all tested RSA modulus sizes, suggesting that the linear regression model provides a reliable approximation of $\varepsilon$.

\begin{figure}[H]
  \centering
  
  \begin{subfigure}[b]{0.45\textwidth}
    \centering
    \includegraphics[width=\textwidth]{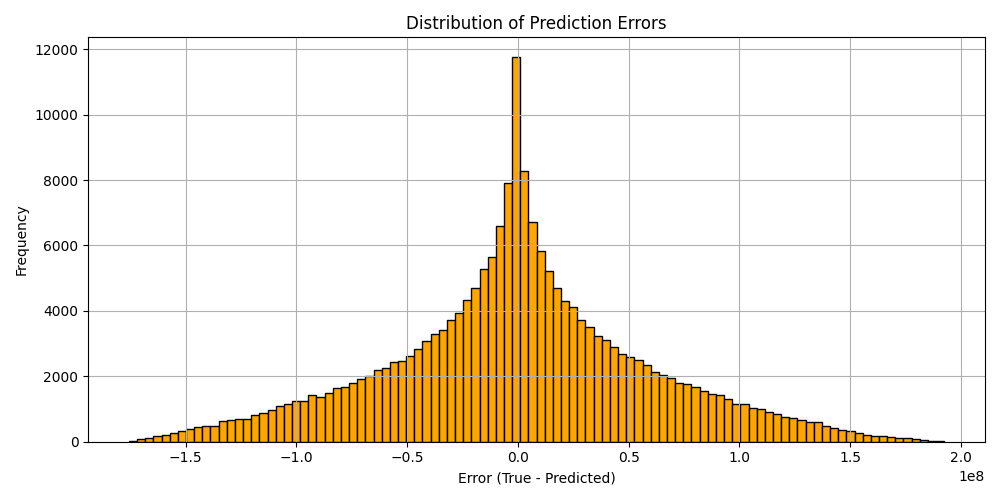}
    \caption{64-bits modulus}
  \end{subfigure}
  \hfill
  \begin{subfigure}[b]{0.45\textwidth}
    \centering
    \includegraphics[width=\textwidth]{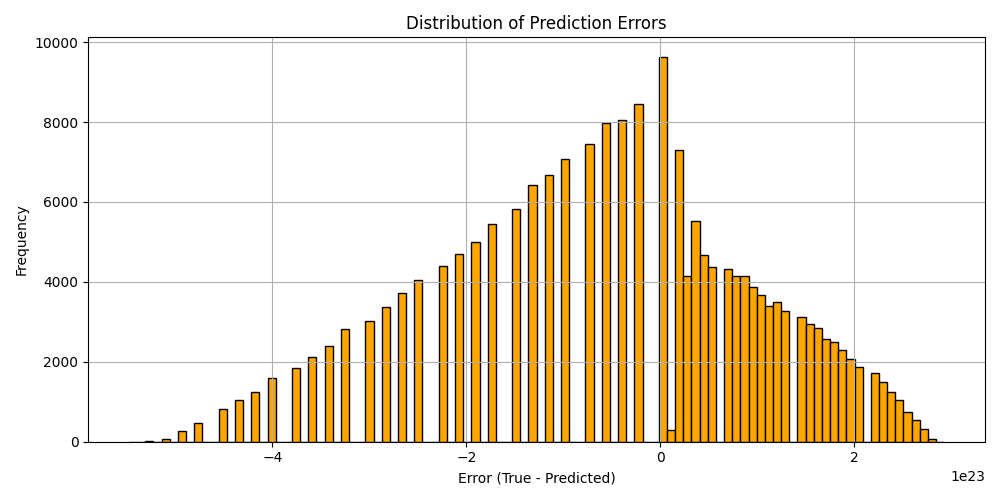}
    \caption{128-bits modulus}
  \end{subfigure}
  
  \vspace{0.5cm}
  
  \begin{subfigure}[b]{0.45\textwidth}
    \centering
    \includegraphics[width=\textwidth]{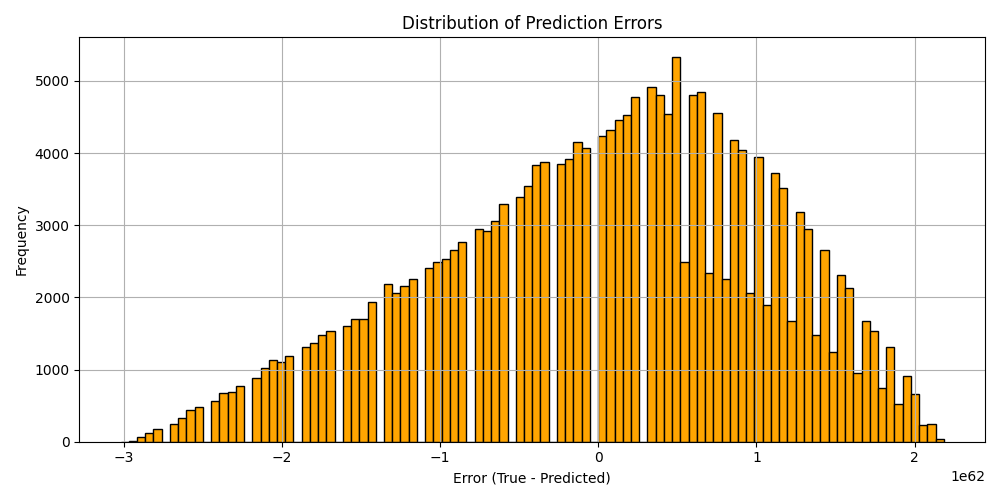}
    \caption{256-bits modulus}
  \end{subfigure}
  \hfill
  \begin{subfigure}[b]{0.45\textwidth}
    \centering
    \includegraphics[width=\textwidth]{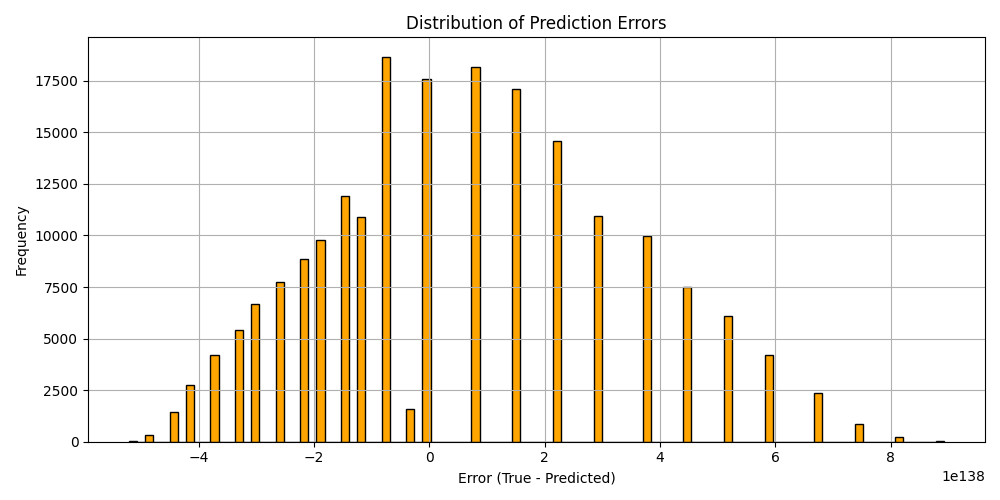}
    \caption{512-bits modulus}
  \end{subfigure}
  
  \caption{distribution of errors}
\end{figure}


\subsection{Discussion}
It follows from our machine learning approach that one should consider custom predictors of size $x$ bits depending on the size of the modulus to be attacked following the algorithm:\\ \\
\begin{algorithm}[H]\label{Predictor generation}
\KwIn{$n$}
\KwOut{$Predictor$}
\textbf{Initialization}: $x \gets number \ of \ bits \ of \ n$  \\
Generate a dataset of $x$ bits RSA modulus with known factors with their $\varepsilon$ values\\
Train a model to build a predictor that best predicts $\varepsilon$\\
\Return $predictor$
\caption{{\sc Predictor} : }
\end{algorithm}

\section*{Conclusion}
Our investigation demonstrates that machine learning, particularly linear regression, can effectively approximate Euler's totient function $\phi(n)$ for RSA moduli across varying bit lengths with a small relative error margin. While deterministic computation of $\phi(n)$ remains infeasible for cryptographically secure parameters, our findings suggest that approximation-based techniques could serve as a supplementary tool in cryptanalysis, potentially weakening RSA's security assurances under certain conditions. This work underscores the growing intersection between statistical learning and cryptanalysis, highlighting the need for further research into the robustness of cryptographic primitives against machine learning-driven attacks. Future directions include exploring more sophisticated models, adversarial training scenarios, and quantifying the practical implications of such approximations in real-world cryptosystem breaches. These insights contribute to the broader discourse on post-quantum and learning-assisted cryptanalysis, urging a re-evaluation of classical hardness assumptions in the era of advanced computational methods.

\bibliography{sample}
\end{document}